\newcommand*{\chn}{}
\newcommand*{\ch}{}

\documentclass[twocolumn]{aastex62}

\usepackage{savesym}
\usepackage{amsmath,amstext}
\usepackage{gensymb}
\let\splitbox\relax 
\usepackage{blindtext}
\usepackage{rotating}
\usepackage[caption=false]{subfig}

\def\C18O{C$^{18}$O}
\def\3CO{$^{13}$CO}
\def\2CO{$^{12}$CO}

\def\N2Dp{N$_2$D$^+$}

\graphicspath{{./}{figures/}}

\submitjournal{AAS Journals}

\shorttitle{Sampling Interstellar Objects}
\shortauthors{Hsieh et al.}

\begin{document}

\title{Evidence Suggesting that 'Oumuamua is the $\sim30\,$Myr-old product of a Molecular Cloud}

\correspondingauthor{Cheng-Han Hsieh}
\email{cheng-han.hsieh@yale.edu}

\author[0000-0003-2803-6358]{Cheng-Han Hsieh}
\affiliation{Department of Astronomy, Yale University, New Haven, CT 06511, USA}

\author{Gregory Laughlin}
\affiliation{Department of Astronomy, Yale University, New Haven, CT 06511, USA}

\author{H{\'e}ctor G. Arce}
\affiliation{Department of Astronomy, Yale University, New Haven, CT 06511, USA}

\begin{abstract}

The appearance of interstellar objects (ISOs) in the Solar System -- and specifically the arrival of 1I/'Oumuamua -- points to a significant number density of free-floating bodies in the solar neighborhood. We review the details of 'Oumuamua's pre-encounter galactic orbit, which intersected the Solar System at very nearly its maximum vertical and radial excursion relative to the galactic plane. These kinematic features are strongly emblematic of nearby young stellar associations. We obtain an a-priori order-of-magnitude  age estimate for 'Oumuamua by comparing its orbit to the orbits of 50,899 F-type stars drawn from  Gaia DR2; a diffusion model then suggests a $\sim$ 35 Myr dynamical age. We compare 'Oumuamua's orbit with the trajectories of individual nearby moving groups, confirming that its motion is fully consistent with membership in the Carina (CAR) moving group with an age around 30 Myr. 
We conduct Monte Carlo simulations that trace the orbits of test particles ejected from the stars in the Carina association. The simulations indicate that in order to uniformly populate the $\sim10^6$ pc$^3$ volume occupied by CAR members with the inferred number density, $n=0.2\,{\rm AU}^{-3}$, of ISOs implied by Pan-STARRS' detection of 'Oumuamua, the required ejection mass is $M\sim 500$\,$M_{\rm Jup}$ per known star within the CAR association. This suggests that the Pan-STARRS observation is in significant tension with scenarios that posit 'Oumuamua's formation and ejection from a protostellar disk.
\end{abstract}


\section{Introduction} 
\label{sec:intro}

The discoveries of the first two interstellar objects (ISOs), 1I/'Oumuamua and 2I/Borisov provide a study in contrasts. 'Oumuamua displayed a number of startling properties, including a light curve with large variations \citep{2017Natur.552..378M,2017ApJ...850L..36J,2017ApJ...851L..38B,2018ApJ...852L...2B,2017ApJ...851L..31K}, a lack of coma or detectable out-gassing \citep{2018AJ....156..261T}, an anomalous component to its acceleration \citep{2018Natur.559..223M}, and an unusual pre-encounter trajectory that placed it nearly exactly at the local standard of rest \citep{2017RNAAS...1...21M}. The second interstellar object, Borisov, by contrast, has behaved in every respect in the manner expected of comets \citep{2020AJ....160...26B}.

'Oumuamua's composition and its point of origin have been the subject of debate from the very moment it was discovered. Its kinematics suggest that it is very young, and a detailed analysis of its trajectory  strongly suggest kinship with a local moving group.  \citet{2020AJ....159..147H} showed that the 
Carina and Columba Associations provide particularly compelling matches.

The galactic population of free-floating objects is continually augmented by icy planetesimals that are ejected from protoplanetary disks by close orbital encounters with embedded giant planets. The Solar System itself is estimated to have contributed of order $M\sim 30M_{\oplus}$ of planetesimals to interstellar space, largely as a consequence of scattering by Jupiter \citep{2005Natur.435..459T, 2008Icar..196..258L}.
A process of this type was almost certainly the mechanism behind comet Borisov's presence as a freely orbiting object in the galactic potential, and it remains a leading hypothesis for 'Oumuamua's origin, although, as we argue here, this interpretation poses difficulties.

Water ice is a dominant constituent of Solar System comets, but, as noted by \citet{2019arXiv190500935S}, water's large enthalpy of sublimation precludes it from generating 'Oumaumua's observed acceleration. Moreover, the simultaneous lack of coma and presence of non-gravitational acceleration has not been observed among comets. 'Oumuamua's shape, which was shown with $\sim90$\% confidence by \citet{2019MNRAS.489.3003M} to resemble an oblate (6:6:1) spheroid, does not resemble the aspect ratio of known Solar System bodies. Moreover, comets are expected to persist indefinitely in the galactic environment, and so it is surprising that the first-detected ISO had kinematics associated with an extremely young age.

\citet{2018A&A...613A..64F} raised the possibility that 'Oumuamua could be composed of H$_2$ ice. This hypothesis was further developed by \citet{2020ApJ...896L...8S} who showed that solid hydrogen can plausibly explain 'Oumuamua's unusual properties. First, with its low  enthalpy of sublimation ($S\sim1\,{\rm kJ/mol}$), exposed H$_2$ ice need only cover several percent of 'Oumuamua's surface in order to produced the observed acceleration. Any out-gassed flux of molecular H$_2$ would have eluded detection. H$_2$ ice has a limited lifetime in the galactic environment, and \citet{2020arXiv200608088H} show that a pure H$_2$-ice object with the current size of 'Oumuamua will last of order $\tau\sim10\,{\rm Myr}$. The transient nature of  H$_2$ ice thus naturally accounts for 'Oumuamua's exceedingly young kinematics, as well as the strange shape. As pointed out by \citet{2017RNAAS...1...50D}, an object that experiences uniform mass loss from its entire surface will develop a large axis ratio prior to disappearing completely. We stress that the formation of objects whose primary initial component is H$_2$ ice \textit{has not been observed}, although a literature exists which explores how such objects might form in the cold, dense cores of giant molecular clouds (see, e.g., \citet{1996Ap&SS.240...75W}). A recent study by \citet{2021ApJ...912....3L} finds that the requirement of extremely low temperatures presents the primary apparent bottleneck to present-day formation of ISOs with a significant solid-H$_2$ component.

The plan for this paper is as follows. In \S 2 we use 'Oumuamua's observed trajectory in conjunction with a recent model of the galactic potential to review the properties of 'Oumuamua's pre-encounter orbit in the Galaxy. Our dynamical integrations highlight the curious fact that when 'Oumuamua was intercepted, it was very close to the upper limits of both its radial and vertical motions in the galactic disk. In \S 3 we compare 'Oumaumua's pre-encounter kinematics with the motions of nearby young associations. This exercise illustrates the apparent connection to the Carina and Columba Young Associations that was discovered by \citep{2020AJ....159..147H}. In particular, orbital integration of 'Oumaumua demonstrates full consistency with membership in the Carina Association. In \S 4 we use Monte Carlo simulations to demonstrate that, given the relatively small number of stars associated with the Carina moving groups, it is highly unlikely that an object ejected from a disk surrounding one of the Carina stars would appear within the search volume probed by Pan-STARRS. 'Oumuamua's appearance suggests that the GMC that gave rise to the Carina association produced a large quantity of short-lived objects, but even this interpretation is problematic given the vast number of such objects that are inferred.

\begin{figure*}[!htp]
\centering
\subfloat{%
  \includegraphics[width=0.44\textwidth]{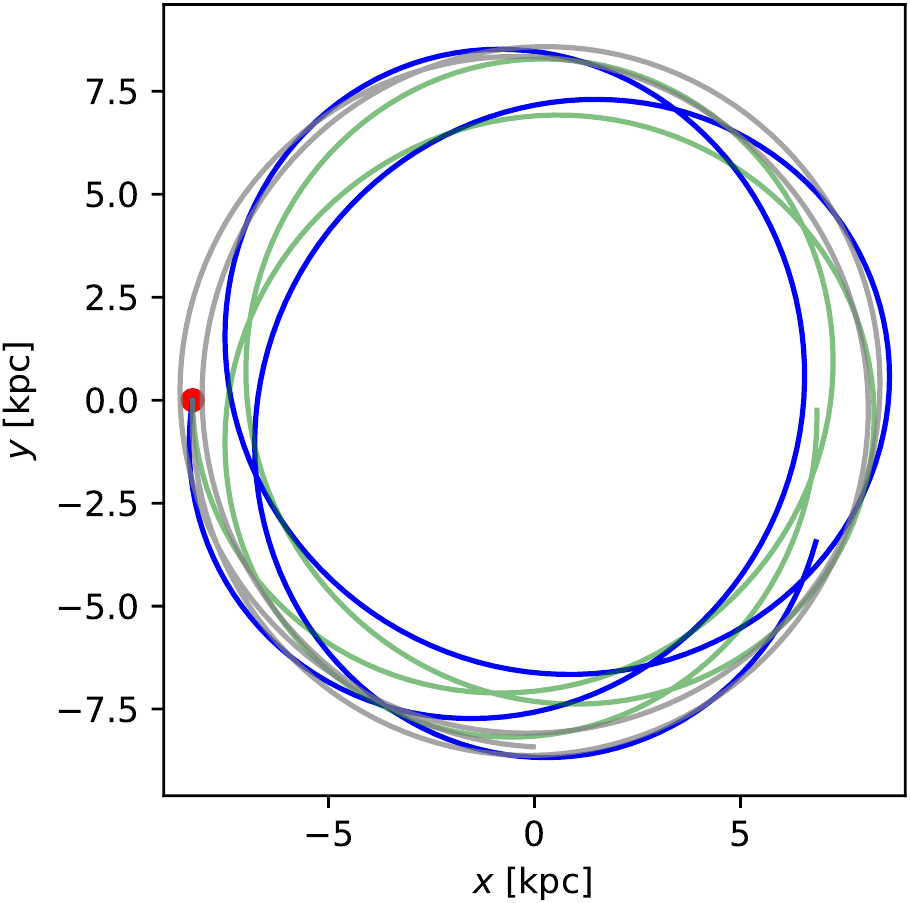}%
}\quad
\subfloat{%
  \includegraphics[width=0.43\textwidth]{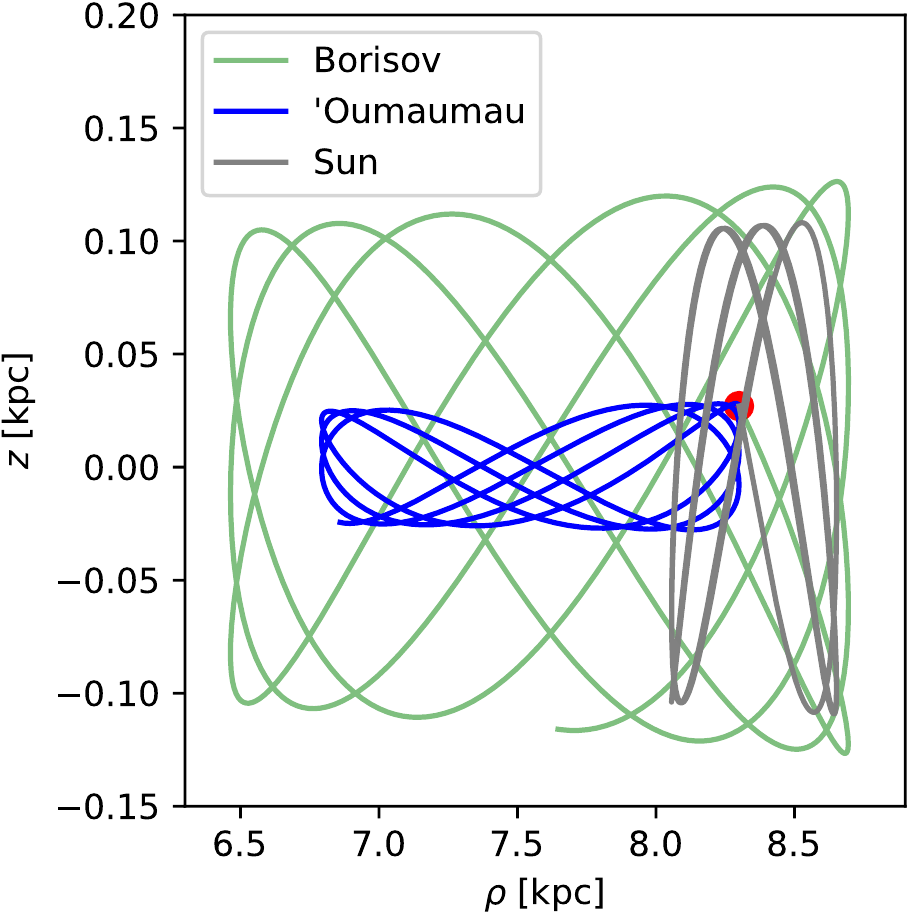}
}
\caption{`Oumuamua and Borisov's galactic orbit integrated for 500 million years into the past. The left plot is in Cartesian coordinates and the right plot in cylindrical coordinates with $\rho$ representing the radial distance from the Galactic Center. The red dot marks the current position of the Sun.}
\label{fig1}
\end{figure*}

\section{`Oumuamua's pre-encounter orbit}
\label{sec1}

We use the Rebound \citep{2012A&A...537A.128R} integration package's high-precision IAS15 scheme \citep{2015MNRAS.446.1424R} to model `Oumaumua Solar System trajectory, as well as Borisov's trajectory for comparison. We adopted ephemerides from NASA's JPL HORIZONS database\footnote{https://ssd.jpl.nasa.gov/horizons.cgi}, and modeled gravitational forces from the Sun and planets in the Solar System  \citep{2015MNRAS.446.1424R}. Non-gravitational forces, which amounted to ${ a}\sim0.001\,{g}_{\odot}\hat{ r}$ for `Oumuamua during the 2-month interval that it was observed  \citep{2018Natur.559..223M} were not included.

We first integrated backward to determine the distance and velocity of the two objects with respect to the Sun a century ago (specifically, we backtraced to December 25th 1919). At that moment, `Oumaumua was moving at near-constant heliocentric velocity of 26.47\,km\,s$^{-1}$ and was at a heliocentric distance of  551.63\,au, whereas Borisov had a heliocentric velocity of 32.33\,km\,s$^{-1}$ and a heliocentric distance of 685.39\,au.

We used HORIZONS to calculate the average change in RA and DEC for 'Oumuamua and Borisov over a ten-year time interval spanning 1919 through 1929, giving the objects' pre-encounter proper motions in the plane of sky with respect to the Sun. Combining the proper motions with the known speeds relative to the Sun, we obtained the pre-encounter heliocentric radial velocities. This information is needed for calculating galactocentric orbits. 

To simulate the orbits within the Milky Way, we first converted the ICRS coordinates employed in the Solar System trajectory calculations to the galactic coordinate system. The velocity of the Sun in galactic coordinates \citep{2010MNRAS.403.1829S}\footnote{As implemented in https://docs.astropy.org/en/stable/coordinates/} is ($U_{\odot}$=11.1, $V_{\odot}$=232.24, $W_{\odot}$=7.25) km\,s$^{-1}$ and the vertical position of the Sun is $z=27.0$ pc above the galactic disk mid-plane. In this system, in 1919, `Oumuamua was located at (X, Y, Z = {\chn -8299.95490805, 0.00226655, 27.0007947}) pc relative to the galactic center and was moving with a galactic velocity of ({\chn $U_O=$-0.522, $V_O=$209.779, $W_O=$-0.574}) km\,s$^{-1}$.\footnote{Note that we did not keep track of the uncertainties for the X, Y, Z, U, V, W coordinates of 'Oumuamua and Borisov in 1919. The Rebound simulated orbits at solar system scale have extremely small uncertainty as compared to the Galactic simulations by Gala. The uncertainty is also well within the uncertainty of the sun’s X, Y, Z, U, V, W inside the Milky Way.}
Its relative velocity with respect to the Sun was ({\chn$U=$ -11.622,$V=$ -22.461,$W=$ -7.824}) km\,s$^{-1}$ in good agreement with the heliocentric velocity ($U=$-11.457, $V=$-22.395, $W=$-7.746) km\,s$^{-1}$ previously reported by \citet{2017RNAAS...1...21M}. 

Similarly, in 1919, Borisov was located at (X, Y, Z =  -8299.95835444, 0.00242457, 26.99990599) pc relative to the galactic center and was moving with a galactic velocity of ($U_O=33.005$, $V_O=208.486$, $W_O=8.286$) km\,s$^{-1}$, giving a relative velocity with respect to the Sun of ($U=21.905$, $V=-23.754$, $W=1.036$) km\,s$^{-1}$.
When combined with the Solar System's galactocentric position, these velocities permit the bodies' trajectories in the galactic potential to be examined.

We use Gala \citep{2017JOSS....2..388P}, an Astropy-affiliated package, to simulate the galactic orbits of the ISOs. We employ a model of the Milky Way potential that consists of a bulge and nucleus \citep{1990ApJ...356..359H}, a disk \citep{1975PASJ...27..533M}, and an NFW halo \citep{1997ApJ...490..493N}. These structural parameters were chosen for consistency with \citet{2015ApJS..216...29B}. We used a high-order Dormand-Prince 85(3) integration scheme \citep{1978JCoAM..18..223D} to trace the orbits of `Oumuamua, Borisov and 796,757 known solar neighborhood stars for 500 million years into the past. All of the bodies were followed with a time-step resolution of one million years. 

The simulated orbits (in the $x$-$y$ plane and in the $\rho$-$z$ plane) for `Oumuamua, Borisov, and the Sun are shown in \autoref{fig1}, which emphasizes a striking feature of `Oumuamua's orbit. It encountered the Sun at a moment where it was close to simultaneously experiencing both its maximum radial and vertical excursions. At the moment of encounter, it had z\,$\sim$\,27\,pc and its current normalized radial position was ${(R_{0}-R_{min})}/{\Delta R}=0.965$), where $R_{0}$ is the current position, $R_{min}$ is the minimum radial excursion from the Galactic Center and the $\Delta R$ is the radial excursion range. Our integration indicates that `Oumuamua spends less than $\sim 2\,\%$ of its time with z\,$\ge$\,27\,pc and $({R_{0}-R_{min}})/{\Delta R}>0.965$. This coincidence would generally be unexpected for a first detection, and suggests that 'Oumuamua belongs to a short-lived population.

\section{'Oumuamua's age and point of origin}

The orbits of interstellar objects correlate with age. For the specific case of 'Oumaumua, \citet{2018MNRAS.480.4903A} investigated how the orbital eccentricity, $e$, the maximum excursion from the galactic plane, $z_{max}$, the perigalactic radius, $R_{min}$, and the apogalactic radius, $R_{max}$ of stars evolve with time, and they used these relations to derive a kinematic age for `Oumuamua of  $\tau_{\rm `O}\sim0.5\,$Gyr, indicating that the ISO is quite young in comparison to the Galaxy.

\citet{2018MNRAS.480.4903A}'s age estimate draws on the correlation between the galactic orbital parameters and the isochrone ages of 14,139 stars in the Geneva-Copenhagen Survey \citep[GCS;][]{2004A&A...418..989N}. In the GCS, however, more than 90\,\% of the sample stars are older than 1 Gyr, so the statistics are sparse for characterizing the generic orbital properties of the youngest stars. The phase space distribution function for ISOs of size $D\ge100\,{\rm m}$, by comparison, is quadrillions of times more finely grained than that of stars, and this contrast motivates an independent estimate of 'Oumuamua's (and Borisov's) ages.
 
We do this by comparing the orbits of the interstellar objects to the orbits of high-mass and low-mass stars in the solar vicinity. High-mass stars persist for a few Gyr, while many low-mass stars do not evolve significantly in a Hubble time. On average, the vertical dispersion increases with age, permitting a kinematic age estimate for an ISO with a known orbit. 

\begin{figure}
\centering
\includegraphics[width=\hsize]{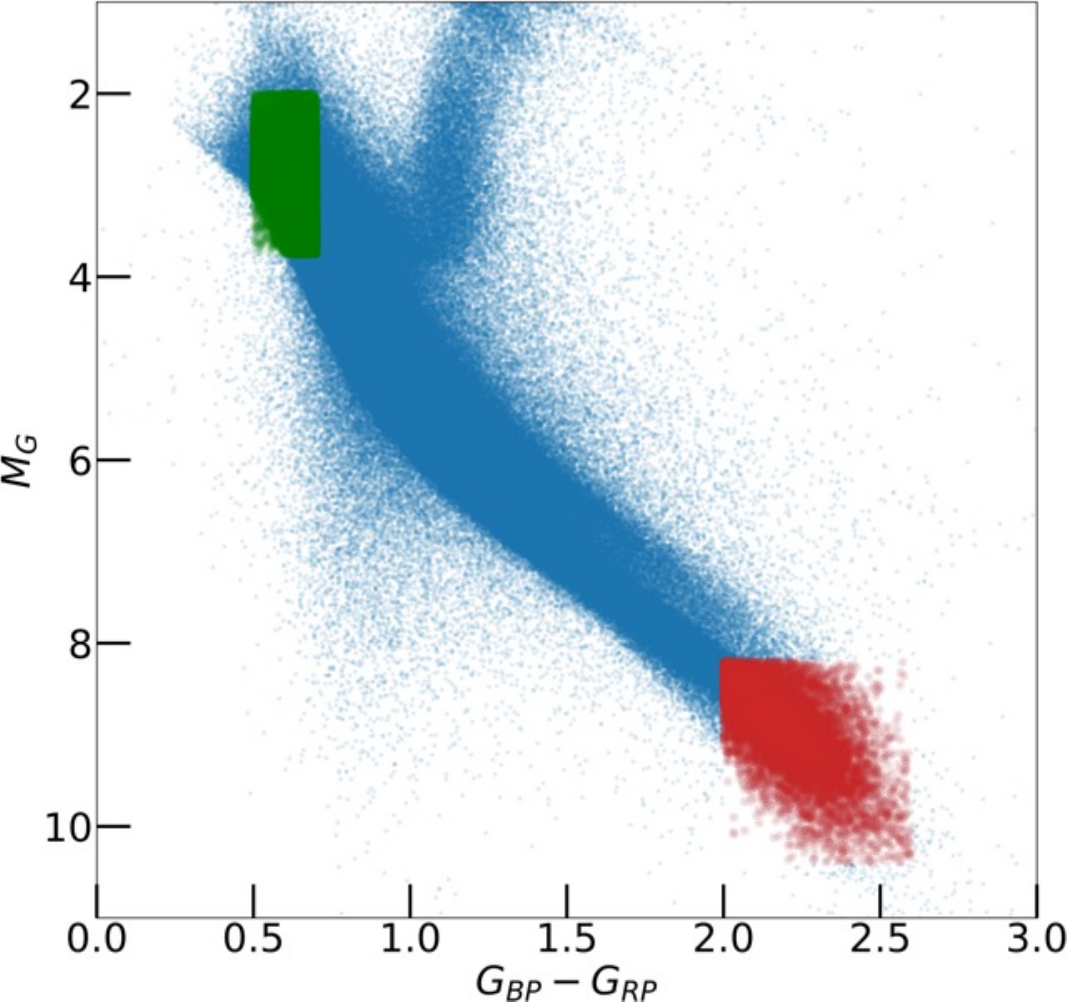}
\caption{Gaia color-magnitude diagram for 796,757 stars within 300 pc of the sun. The selected 50,899 high-mass stars and 13,066 low-mass stars are highlighted in green and red respectively.}
\label{fig2}
\end{figure}

We first select the 796,757 stars from the DR2 catalog \citep{2018A&A...616A...1G} having $d<300\,{\rm pc}$, along with complete radial velocity and proper motion information. Proceeding from the color-magnitude diagram of these stars, we apply the following criteria to separate high-mass (HM) {\chn (typically F-type stars)} and low-mass (LM) stars {\chn (typically K-\,\&\,M-type dwarfs)}:
  \begin{gather}
      HM : 
      0.5 < G_{BP}-G_{RP} < 0.7 \\
      2 < M_G < 3.75
   \end{gather}
    \begin{gather}
    LM : 2.0 < G_{BP}-G_{RP} < 2.6\\
      8.2 < M_G < 10.4{,}
   \end{gather}
where $G_{BP}$ and $G_{RP}$ are the Gaia bands covering (330--680 nm) and (630--1050 nm) in wavelengths respectively.
The two groups are shown in \autoref{fig2}.  We restrict the analysis to stars located within 200\,pc of the mid-plane to prevent biasing the sample toward high-mass stars as a consequence of their intrinsic luminosities. This cut-off falls within the $z\sim300\,{\rm pc}$ stellar disk scale height \citep{1991ApJ...378..131K, 2002A&A...394..883L,2011MNRAS.414.2446M}.

For each retained star, we integrate the orbit in the galactic potential (using the procedure described previously) to determine the star's maximum vertical excursion. 
For an individual star, ${z}_{\rm max}$ can be approximated by a random walk process, so that ${z}_{\rm max}\propto \sqrt{t}$. Knowing that 'Oumaumua has ${z}_{\rm max}=27.97\,$pc, and that  Borisov has ${z}_{\rm max}=126.46\,$pc, the simple diffusion approximation provides an estimate of their ages.

The ${z}_{\rm max}$ distribution of K-\& M-type dwarfs has a long tail toward higher values, providing deviation from a Gaussian distribution. The average ${z}_{\rm max}$ for the low-mass sample is $\sim260$ pc. This is slightly larger than 210\,pc ${z}_{\rm max}$ for the higher-mass (F-type) stars. The random scattering process follows ${z}_{\rm max} \propto t^{0.5}$, which has the rate of dispersion flattening with time. The small ${z}_{\rm max}$ difference between the low-mass sample and the high-mass sample indicates ${z}_{\rm max}$ is not strongly sensitive to residence times ranging from 2\,Gyr to $\gtrsim 10\,{\rm Gyr}$ (the approximate age of the galactic disk). `Oumuamua, in particular, has a ${z}_{\rm max}$ only 1/10th of the average ${z}_{\rm max}$  for the low-mass sample, placing it in the region where ${z}_{\rm max}$ is sensitive to age.  We note that in general, with the ${z}_{\rm max} \propto t^{0.5}$ approach, if we compared `Oumuamua’s ${z}_{\rm max}$ to the ${z}_{\rm max}$ of higher mass objects, we would get a more accurate age estimate. We do not use O, B, and A stars on account of their paucity in the Solar neighborhood. F-type stars with a Gaussian-like distribution in ${z}_{\rm max}$, present the best compromise between youth and abundance.  

The F-type stars satisfying criteria 1 and 2 have an average age, $\bar{\tau}_{F}\sim 2$ Gyr \footnote{F-type main sequence stars have maximum ages of about 4 to 5 billion years \citep{2013ApJ...771...40B}. For a uniform sampling of F-type stars, a very rough average age should be around 4/2 = 2 billion years.}, during which their vertical excursions have evolved to $\bar{z}_{\rm F}=$210 $\pm 160$ pc. We can thus estimate the age, $\tau_{\rm ISO}$ of an ISO using ${z_{\rm ISO}}/{\bar{z}_{\rm F}}=\sqrt{{\tau_{\rm ISO}}/{\bar{\tau}_{\rm F}}}$. The vertical excursions of the orbits are obtained from the means of the maximum distances above and below the disk. The maximum vertical excursion of {\chn 27.97}\,pc that we find for 'Oumuamua agrees with the $27.71\,$pc value calculated by \citet{2018MNRAS.480.4903A} (See \citet{2018MNRAS.480.4903A} Table 1 for comparison)\footnote{The calculation assumes the Sun is located at 27\,pc above the Galactic plane. `Oumuamua is at its maximum vertical excursions, and the maximum vertical excursion would be close to the current height of the sun from the Galactic mid-plane.}.

The diffusion models thus gives a $\tau_{\rm `O}\sim\,35\,$Myr age for 'Oumuamua and a $\tau_{\rm B}\sim$\,710\,Myr age for Borisov.
While rough, this estimate suggests that 'Ouamuamua may be significantly younger than 500\,Myr. It is thus worthwhile to establish whether it can be connected to known nearby young stellar associations.

\begin{figure*}[tbh!]
\centering
\makebox[\textwidth]{\includegraphics[width=\textwidth]{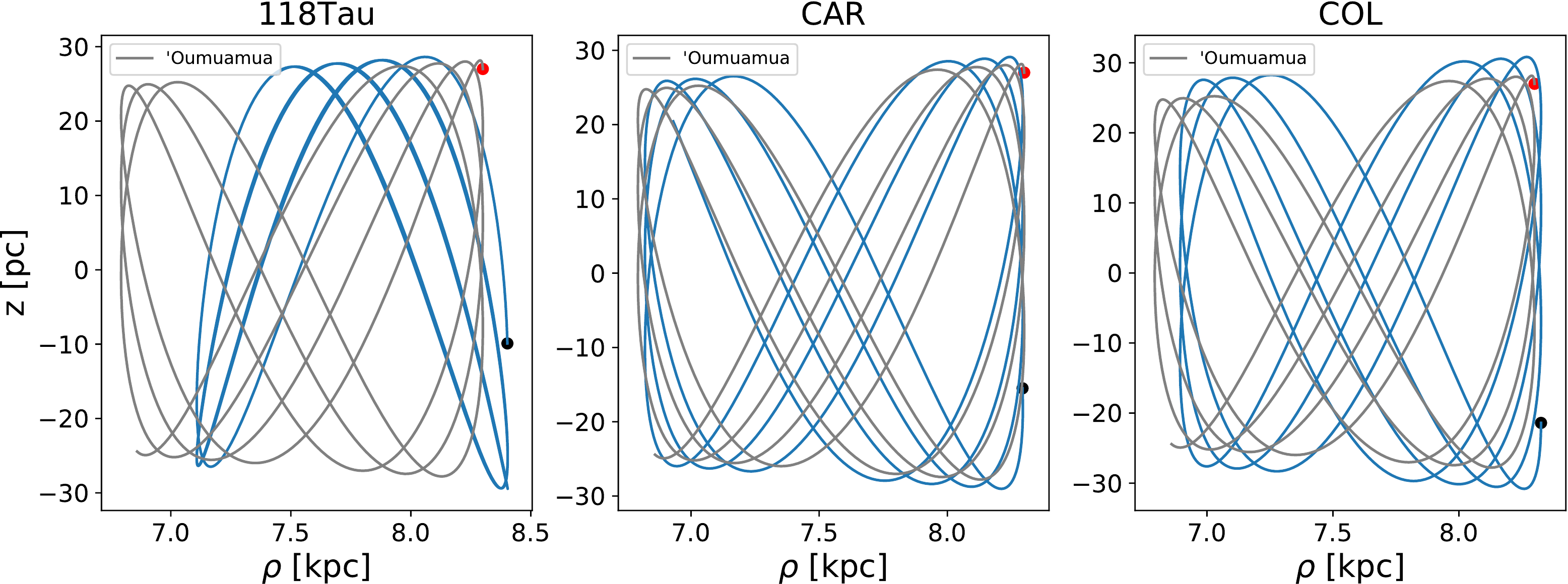}}
\caption{The orbit of `Oumuamua in the Galactic potential compared to the orbits of three young associations. In each panel, trajectories are propagated 500 million years into the past in order to facilitate comparison. The quantity $\rho$ charts the distance from the Galactic Center. The red and black dots mark the current position of the Sun and the corresponding young association, respectively. }
\label{fig6}
\end{figure*}

Work by \citet{2018ApJ...856...23G} draws on the Gaia DR1 to identify 27 young stellar associations within 150\,pc of the Sun, and tabulates the averaged Galactic positions ($X$, $Y$, $Z$) and velocities ($U$, $V$, $W$) for the associations. Using the procedure described above, we integrate these centroids backward in the galactic potential. The results for the Carina (CAR), Columbia (COL) and 118 Tau moving groups are plotted in \autoref{fig6} and the key properties derived from the simulations of all 27 young stellar associations are shown in \autoref{table:1}.

\begin{table*}[t]
\setlength{\tabcolsep}{3.5pt} 
\caption{Summary of the simulated orbits of the 27 young stellar associations within 150\,pc}             

\label{table:1}      
\centering                          
\begin{tabular}{c c c c c c c c c c c c c c c}        
\hline\hline                 

&118Tau & ABDMG & BPMG & CAR & CARN & CBER & COL & CRA & EPSC & ETAC & HYA & IC2391 & IC2602  &LCC \\
\hline\hline
$R_{max}$\tablenotemark{a}&8.4&8.31&8.3&8.29&8.39&8.43&8.33&8.2&8.26&8.27&8.76&8.34&8.27&8.26\\
\hline
$R_{min}$\tablenotemark{a}&7.11&6.48&7.21&6.82&7.01&7.87&6.89&7.08&6.95&6.77&6.78&7.27&6.85&6.79\\
\hline
$z_{max}$\tablenotemark{a}&29.45&99.4&29.09&29.11&70.46&129.59&30.84&47.79&42.78&71.76&86.75&30.85&94.82&15.76\\
\hline
$R_{0}$\tablenotemark{a}&8.4&8.31&8.3&8.29&8.3&8.31&8.33&8.17&8.25&8.27&8.34&8.3&8.25&8.25\\
\hline
$Z_{0}$\tablenotemark{a}&-9.9&-8.8&-15.7&-15.5&-4.3&84.9&-21.4&-42.43&-25.6&-34.81&-15.8&-18.0&-12.6&5.8\\
\hline
$\frac{R_{0}-R_{min}}{DR}$&1.0&1.0&1.0&1.0&0.93&0.79&1.0&0.97&0.99&1.0&0.79&0.96&0.99&0.99\\[1.2ex]
\hline
$Z_{0}/z_{max}$&0.34&0.09&0.54&0.53&0.06&0.66&0.69&0.89&0.6&0.49&0.18&0.58&0.13&0.37\\
\hline
Age\tablenotemark{b}& 10& 149 & 24 & 25$\sim$30& 200& 562& 42& 4-5 & 3.7 & 11 & 750 & 50 & 46 & 15\\
\hline
Age Ref.&1& 2& 2 & 3 &4 &5 &2 &6 & 7 & 2 & 8 &9 & 10 &11 \\
\hline
& &  &  & & & & &  &  &  & &  &  & \\[3ex]

\hline\hline
&OCT & PL8 & PLE & ROPH & TAU & THA & THOR & TWA & UCL & UCRA & UMA & USCO & XFOR &\\
\hline \hline     
$R_{max}$\tablenotemark{a}&8.3&8.29&8.42&8.2&8.43&8.3&8.39&8.29&8.22&8.19&9.17&8.21&8.33\\
\hline
$R_{min}$\tablenotemark{a}&8.13&6.75&6.55&7.25&7.8&6.89&7.12&7.08&6.85&6.98&7.77&7.2&6.83\\
\hline
$z_{max}$\tablenotemark{a}&71.91&53.19&112.81&38.65&42.06&95.87&34.47&32.38&39.04&40.7&49.0&50.16&85.37\\
\hline
$R_{0}$\tablenotemark{a}&8.3&8.29&8.42&8.18&8.42&8.29&8.39&8.29&8.19&8.16&8.31&8.18&8.33\\
\hline
$Z_{0}$\tablenotemark{a}&-59.7&-13.9&-54.4&37.6&-35.9&-36.1&-23.9&22.7&26.5&-39.2&21.9&48.9&-84.2\\
\hline
$\frac{R_{0}-R_{min}}{DR}$&1.0&1.0&1.0&0.98&0.98&0.99&1.0&1.0&0.98&0.98&0.39&0.97&1.0\\[1.2ex]
\hline
$Z_{0}/z_{max}$&0.83&0.26&0.48&0.97&0.85&0.38&0.69&0.7&0.68&0.96&0.45&0.97&0.99\\
\hline
Age\tablenotemark{b}& 35& 60 & 112 &$<$2 &1-2 &45 & 22& 10 & 16 & 10 &414 & 10 & 500 & \\
\hline
Age Ref.& 12&13 & 14 &15  &16 & 2& 2& 2&  11&  17& 18 &11 & 19 & \\
\hline
\end{tabular}

\begin{flushleft}
\tablerefs{(1) \citet{2016.2016M}, (2)
\citet{2019AJ....157..234S}, (3)
\citet{2015MNRAS.454..593B}, (4) \citet{2006ApJ...649L.115Z}, (5) \citet{2014A&A...566A.132S}, (6) \citet{2012MNRAS.420..986G}, (7) \citet{2013MNRAS.435.1325M}, (8) \citet{2015ApJ...807...24B}, (9) \citet{2004ApJ...614..386B}, (10) \citet{2010MNRAS.409.1002D}, (11) \citep{2016MNRAS.461..794P}, (12) \citet{2015MNRAS.447.1267M}, (13) \citet{1998AJ....116.2423P}, (14) \citet{2015ApJ...813..108D}, (15) \citet{2008hsf2.book..351W}, (16) \citet{1995ApJS..101..117K}, (17) \citet{2018ApJ...856...23G}, (18) \citet{2015AAS...22511203J}, (19) \citet{2010A&A...514A..81P}}
\tablenotetext{a}{In units of pc.}\tablenotetext{b}{In units of Myr.}
\end{flushleft}
\end{table*}

\autoref{table:1} and \autoref{fig6} show that nearly all of the young associations display maximum vertical excursions that are similar to `Oumaumua's. The small values for $z_{\rm max}$ reflect conditions of formation, prior to scattering encounters with molecular clouds and spiral arms. In \autoref{table:1} we also list the current normalized radial positions, $({R_0-R_{min}})/{\Delta R}$. Nearly all of the young associations, like `Oumuamua, have ${R_0-R_{min}}/{\Delta R}\sim 1$. This indicates they are all currently near the apocenters of their Galactic orbits, reflecting recent formation (with respect to the $\sim$250\,Myr Galactic orbital period) with angular velocities that are less than the local circular velocity. 

This phenomenon mainly stems from trapping caused by the gravitational potential perturbation associated with spiral arms, and was studied by \citet{1987ApJ...314...10R}. Numerical simulations by those authors showed that a spiral arm creates a retrograde relative motion that can entrain the material constituting a given cloud complex for $\sim$50 Myr, a time scale comparable to the lifetime of a molecular cloud. In general, molecular clouds form stars when approaching the minimum of the spiral potential. Therefore, as a consequence of the retrograde motion, one concludes that molecular clouds (with lifetimes $<$ 50 Myr) should lie near the apocenters of their galactic orbits. 

Recent simulations by \citet{2018MNRAS.474.2028R} found that gas in the spiral arms can typically have a net radial streaming motion of v$_R \approx -9$ km\,s$^{-1}$ with an azimuthal velocity deficit of order 6 km\,s$^{-1}$ slower than the local circular velocity. This translates to average peculiar motions trending towards the galactic center and against the sense of galactic rotation. (Note that the radial oscillation with periodic modulation from spiral arms has a time scale of $\sim$ 450 Myr \citep{1987ApJ...314...10R}). We thus expect young associations with ages $\ll$ 450 Myr to be near their apocenters and moving with sub-circular azimuthal velocities, which is in excellent agreement with the results in Table 1, and to our knowledge, this is the first direct observational evidence from Gaia showing that stars form at the apocenter of sub-circular orbits.


Among the 27 young associations identified by \citet{2018ApJ...856...23G}, the 30 Myr-old group ``CAR'' \citep{2019AJ....157..234S} and the 42 Myr-old group ``COL'' stand out in terms of their kinematic similarity to `Oumuamua (\autoref{fig6}). Both associations share almost identical maximum vertical excursions, and maximum and minimum radial positions with 'Oumuamua, and are currently located close to their maximum radial positions from the Galactic Center. 

If a true physical association exists, we would expect `Oumuamua's trajectory to intersect the trajectory of the source association at the time of the latter's formation. Moreover, this constraint should be satisfied regardless of whether `Oumuamua was a comet-like object ejected from a protoplanetary disk, or alternately, a product of the parent GMC itself. 

To probe for a physical association, we used the Gala package to explore the range of possible orbits of all 27 young associations. For each association, we integrated 2000 ($X$, $Y$, $Z$, $U$, $V$, $W$) sets drawn from Gaussian distributions in each quantity and conforming to the covariance matrices reported by \citet{2018ApJ...856...23G}. The integrations were all run backward in time for 50 million years with time step of 0.1 million years. 

The effect of disk heating on `Oumuamua's time-reversed trajectory is negligible on a 35 Myr timescale. In the vicinity of the solar neighborhood, the relevant disk heating mechanisms are isotropic heating due to stochastic scattering from molecular clouds and density wave scattering. The density wave mechanism is dominated by the two-arm spiral which encounters stars twice per orbit (on the time scale of $\sim$\,125\,Myr). Since the density wave’s frequency is close to the local epicyclic frequency, the radial random motion is greatly increased by the near resonance. This would result in the radial velocity dispersion ($\sigma_R$) being much greater than vertical velocity dispersion ($\sigma_Z$) \citep{1990MNRAS.245..305J}. Observations have shown that for solar neighborhood stars, $\sigma_Z / \sigma_R\sim0.5$, suggesting that both spiral density waves and scattering by molecular clouds contribute significantly to the heating of the local disk \citep{1998MNRAS.298..387D,2002MNRAS.337..731H}. \citet{1993MNRAS.263..875I} and \citet{2008ASPC..396..341S}, however, suggest that this anisotropy can be explained by Giant Molecular Cloud scattering alone. The time-reversed trajectory for a 35\,Myr run-out unfolds substantially faster than the timescale for spiral-arm heating ($\sim 125$\,Myr). 'Oumuamua's inferred age is also comparable to the lifetime of a typical molecular cloud, further diminishing the effect of molecular cloud scattering.

The main uncertainty in 'Oumuamua's trajectory stems from the uncertainty in the Sun's Galactic position and velocity. We simulate the allowed range of `Oumuamua's time-reversed trajectory by perturbing its nominal orbit with 1000 realizations from the Gaussian uncertainty distributions in (X, Y, Z, U, V, W) associated with the Sun's galactic position and velocity.

In Table 3 of \citet{2017MNRAS.465..472K}, the authors collected more than 50 estimates of the Sun’s height, $Z$, above the Galactic mid-plane. We adopted the median value $Z=17\,$pc as the vertical position of the Sun and set the standard deviation to be $\delta z=10\,$pc, which includes almost all the values listed in their table. For the Sun's radial position, $X$, radial velocity, $U$, and angular velocity, $V$, we adopted $X=8.27\,\pm\,0.29$ kpc, $U=13.84\,\pm\,0.27$ km\,s$^{-1}$ and $V=250\,\pm\,9$ km\,s$^{-1}$ respectively from \citet{2010MNRAS.403.1829S}. For the vertical velocity, $W$, we adopted the value $W=7.3$\, km\,s$^{-1}$ used in \citet{2011ApJS..195...26A}. As noted by \citet{2010MNRAS.403.1829S}, there is a $\sim 4\,$ km\,s$^{-1}$ systematic difference between the average $W$ motion towards the Galactic North and South pole, possibly indicating a systematic error around $\sim 2$ km\,s$^{-1}$. For simplicity, we adopted $\delta W \sim 2$ km\,s$^{-1}$ as the standard deviation of the vertical velocity.

\begin{figure*}[tbhp]
\centering
\makebox[\textwidth]{\includegraphics[width=0.9\textwidth]{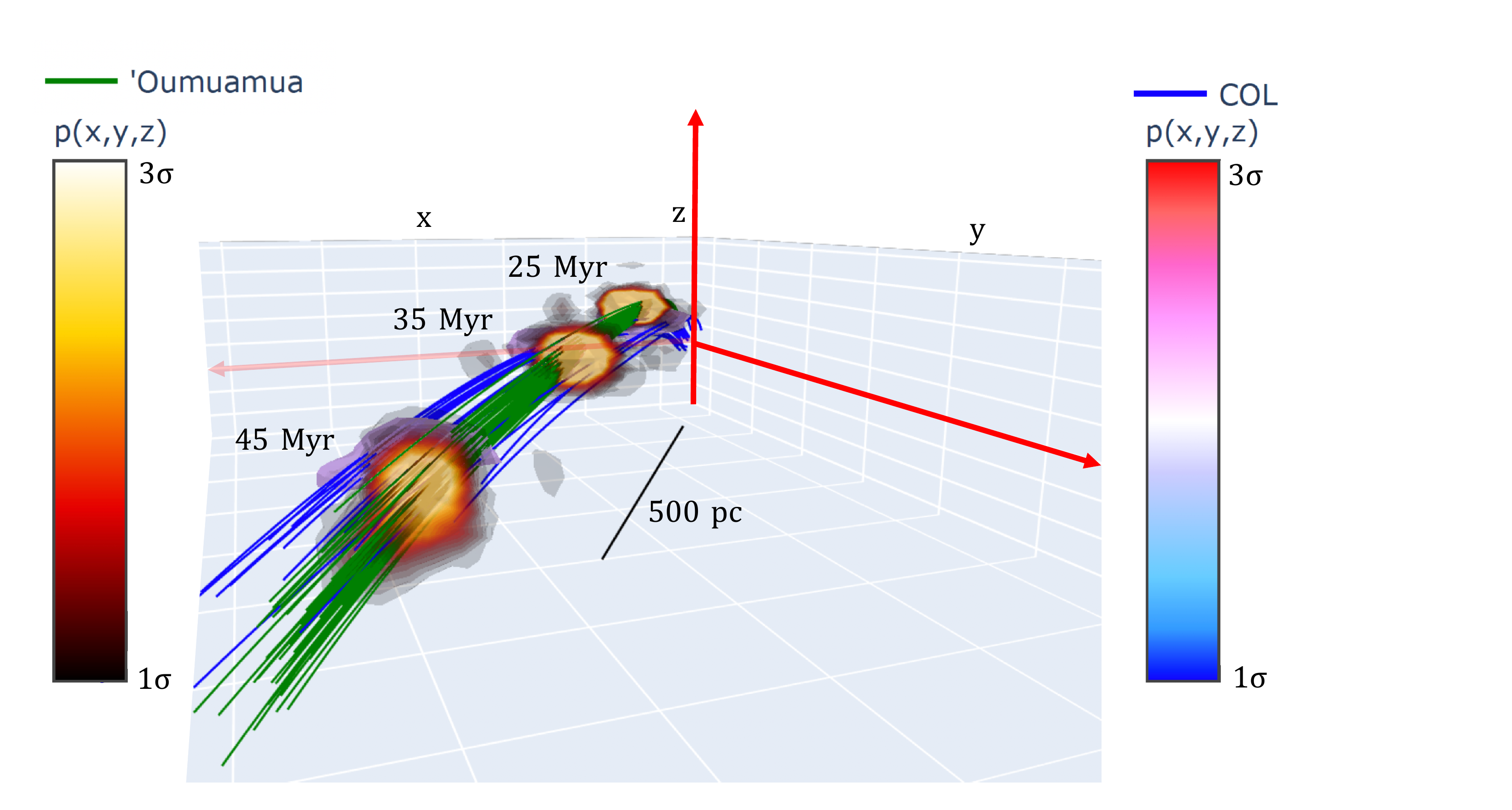}}
\caption{The three-dimensional galactic orbits of `Oumuamua and the COL young stellar association integrated backwards for 50 million years. Positive $x$ is in the direction towards the Galactic Center. Positive $y$ is the Galactic rotation direction. Sample trajectories for `Oumuamua and for stellar members of COL identified by \citet{2018ApJ...856...23G} are represented by the bundles of green and blue lines respectively. The colored volumes mark the 1 $\sigma$ to 3 $\sigma$ surfaces for both Oumuamua and COL at 25 Myr, 35 Myr and 45 Myr. An interactive version of this figure can be found in the online version of the paper.
}
\label{fig8}
\end{figure*}

\begin{figure*}[tbhp]
\centering
\makebox[\textwidth]{\includegraphics[width=0.9\textwidth]{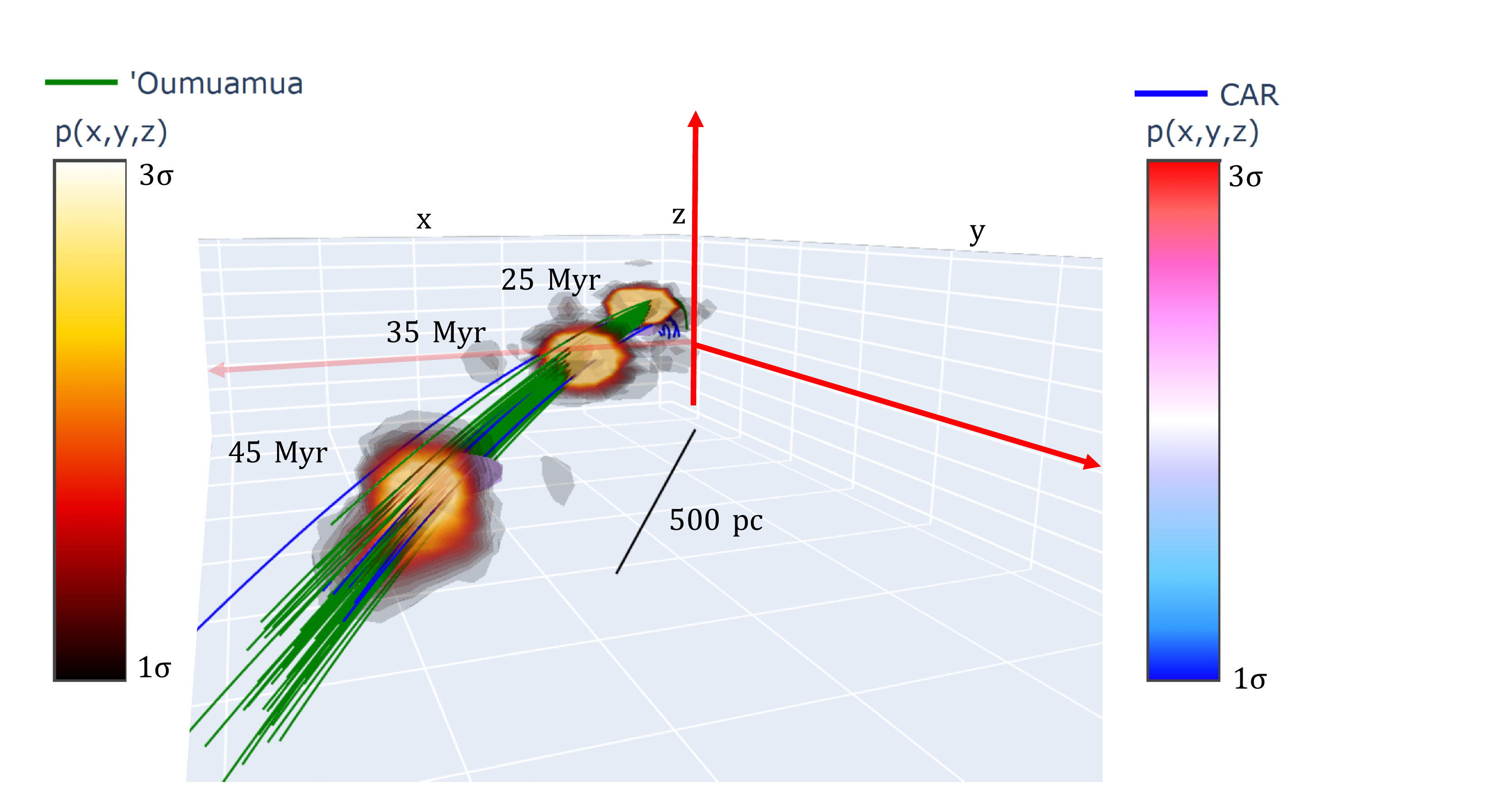}}
\caption{The three-dimensional Galactic orbits of `Oumuamua and the CAR young stellar association integrated backwards for 50 million years. Positive $x$ is in the direction towards the Galactic Center. Positive $y$ is the Galactic rotation direction. Sampled trajectories for `Oumuamua and for the stellar members of CAR identified by \citet{2018ApJ...856...23G} are represented by bundles of green and blue lines respectively. The colored volumes mark the 1-$\sigma$ to 3-$\sigma$ surfaces for both Oumuamua and CAR at 25 Myr, 35 Myr and 45 Myr. An interactive version of this figure can be found in the online version of the paper. 
}
\label{fig9}
\end{figure*}

The simulated 3D galactic orbits for CAR, COL, and 'Oumuamua in the Solar System's frame are shown in \autoref{fig8} and \autoref{fig9}. An interactive version of this figure is in the online version of the paper. Positive x is in the direction towards the Galactic Center. Positive y is in the direction of Galactic rotation. `Oumuamua and COL's trajectories are represented by green and blue lines respectively. The colored volume marks the 1-$\sigma$ to 3-$\sigma$ probability surfaces for both `Oumuamua and COL at 25 Myr, 35 Myr and 45 Myr. Note that the dispersion in position for 1000 clones of `Oumuamua at -50\,Myr are ($\sigma_x$, $\sigma_y$, $\sigma_z$) = (18.9, 29.6, 22.2) pc, or  $\sigma_r$ = 41.5 pc.

Over the last $\sim$50\,Myr CAR and COL have effectively matched orbits with `Oumuamua. To quantify the degree of intersection between `Oumuamua and the young associations, we calculate the 1-$\sigma$ intersection volume between `Oumuamua and the stellar moving groups. The calculation is done by first applying kernel density estimation (KDE) to evaluate the probability of `Oumuamua and an association at each volume element. Then, three-dimensional 1-$\sigma$ masks enclosing 68\% of the probability are calculated to find the intersection. We calculated the 1-$\sigma$ intersection volume for all 27 nearby young stellar associations currently within 150 pc from the Sun between 15 and 50 Myr ago. In total there are only 6 associations that have non-zero 1-$\sigma$ intersection volumes with `Oumuamua.

Of these 6 young associations, only CAR and COL intersect with `Oumuamua's trajectory at their corresponding ages of 30 and 42 Myr. In particular, the CAR association appears to move in concert with `Oumuamua and maximizes its 1-$\sigma$ intersection volume with the ISO at around 34\,Myr. The 34\,Myr peak of CAR's intersection volume is in agreement with the $\sim$25-30 Myr old age based on lithium depletion method and its color magnitude diagram \citep{2019AJ....157..234S}. COL while also intersecting with the ISO, does not maximize its 1-$\sigma$ intersection volume at 42\,Myr. This suggests that `Oumuamua  more likely originated from CAR than from COL. It is also important to point out, while the intersection of 3D Galactic orbits favors the origin from CAR, COL cannot be completely ruled out. The intersections with `Oumuamua at the estimated formation ages, in conjunction with the matching of Galactic orbits constitutes a strong constraint. `Oumuamua can be counted as a member of Carina or the Columba moving groups. This conclusion echoes the work done by \citet{2020AJ....159..147H}, who showed that `Oumuamua passed through the Carina and Columba moving groups at the time when they were forming.

\section{Implications for `Oumuamua's Origin}

\begin{figure*}[tbh]
\centering
\makebox[\textwidth]{\includegraphics[width=0.9\textwidth]{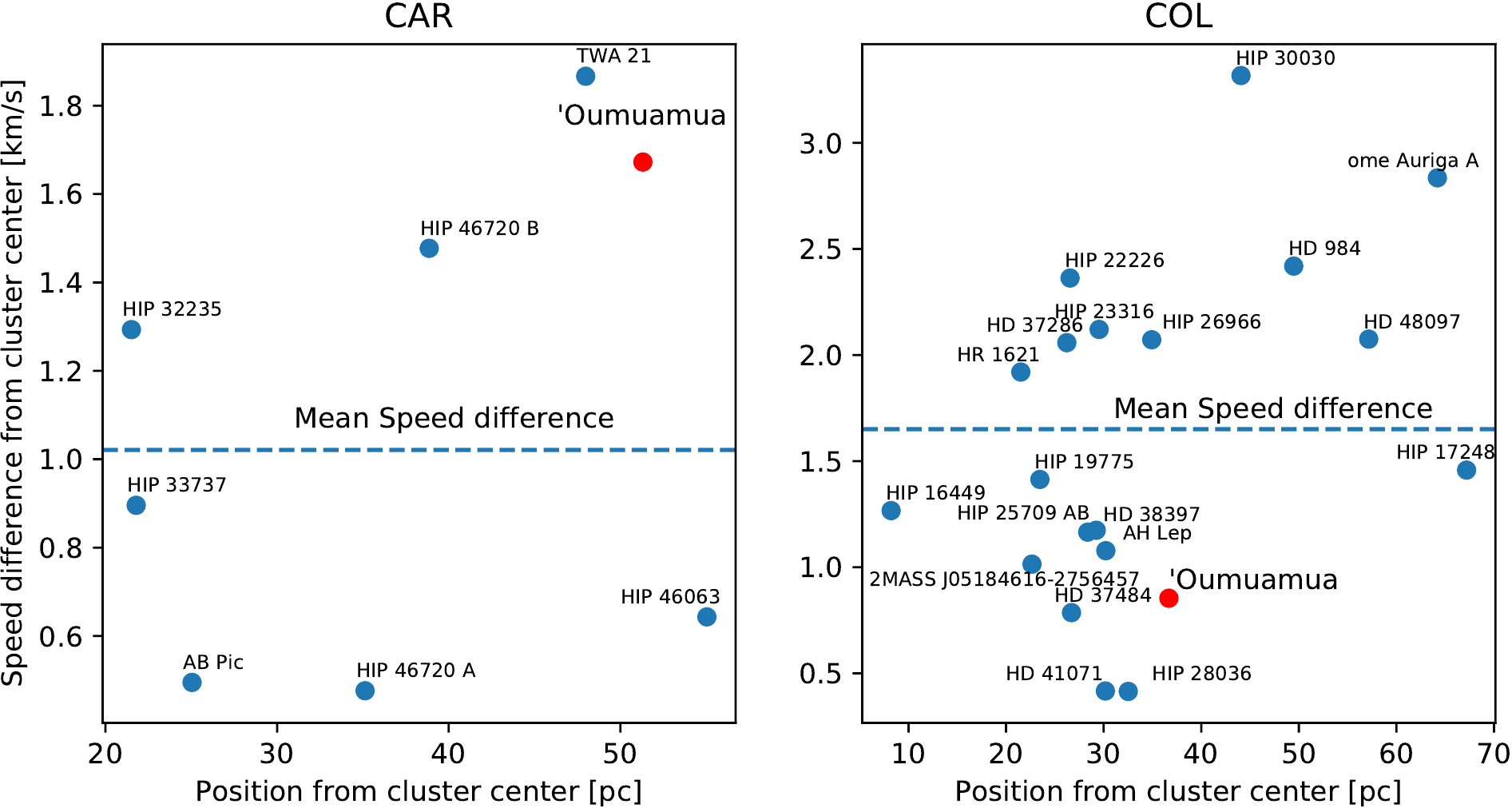}}
\caption{The distributions of COL and CAR member stars in position and velocity space relative to the cluster centers. Note that for `Oumuamua the ejection velocity is plotted on the y-axis instead of the speed difference from the cluster center.  
}
\label{fig11}
\end{figure*}

By calculating Pan-STARRS' aggregated sensitivity, \citet{2018ApJ...855L..10D} estimated that the interstellar number density of `Oumuamua-like objects is of order $n=0.2\, {\rm au}^{-3}$, adopting a cigar or oblate spheroid shape \citep{2018ApJ...856L..21B} with a $>$6:1 axis ratio \citep{2018ApJ...857L...1M}, effective length smaller than 440 meters \citep{2018AJ....156..261T}, (or effective spherical radius of 102\,m) and density of $\sim 3$g\,cm$^{-3}$,  giving a total mass density of 4 $M_{\Earth}$\,pc$^{-3}$. In the previous section, we found support for an origin of `Oumuamua in the Carina or Columba  moving groups. As a consequence, its detection was either a fluke, or we can conclude that similar objects from Carina and Columba suffuse the local interstellar medium. 

\autoref{fig11} charts the
current positions and velocities of the COL and CAR member stars with respect to the centroids of their respective associations. For `Oumuamua we plot its ejection velocity from the two Young Associations and its relative distance from the cluster center. The ejection velocity is estimated by dividing `Oumuamua's current distance from the cluster center with the cluster age listed in \autoref{table:1}. We obtain ejection speeds of 1.67 km\,s$^{-1}$ and 0.85\,km\,s$^{-1}$ for CAR and COL respectively. We note that the current position of `Oumuamua is closer to the cluster center than the furthest association star for both CAR and COL, which would be expected for a member of either of these associations, and the modest speed is consistent with both the disk ejection and molecular cloud byproduct hypotheses \citep{2017RNAAS...1...13G,2020ApJ...896L...8S}. 

An order-of-magnitude estimate seems to distinguish between the scenarios. Using the \citet{2018ApJ...855L..10D}  estimate of 4 $M_{\Earth}$\,pc$^{-3}$, and spherical volumes defined by the furthest member in the cluster ($\sim 55$\,pc for CAR, $\sim 70$\,pc for COL, see \autoref{fig11}), one finds a combined volume for CAR and COL of $\sim 2.1\times10^{6}$\,pc$^3$. Naively, this implies an enormous total of $\sim 8.5\times10^{6}\,M_{\Earth}$ in `Oumuamua-like objects originating from CAR and COL. 

The assigned stellar memberships of CAR and COL vary slightly with the clustering algorithm that is employed, but the associations contain of order 40 stars, implying $ 2.1 \times 10^5 M_{\Earth}$ ejected per star. While crude, this value greatly exceeds the inferred mass of our Solar System’s planetesimal disk ($\sim 12-65\,M_{\Earth}$ \citep{2013ApJ...768...45N,2014ApJ...792..127R,2017AJ....153..153D}, suggesting a difficulty with the disk ejection picture \citep{2017RNAAS...1...13G}. 

To refine the foregoing estimate, we conducted a Monte Carlo disk ejection simulation from the 7 stars in COL shown in \autoref{fig11}. We first back-traced the stars in \autoref{fig11} with time steps of 0.1\,Myr and recorded their galactic positions 30\,Myr ago. For each star at this origin moment, we launched 50,000 test particles uniformly over the 4$\pi$ range of directions with ejection velocities sampled from a Boltzmann-like distribution
\begin{gather}
f\left(x ; \frac{1}{\beta}\right)=\frac{1}{\beta} \exp \left(-\frac{x}{\beta}\right)\, ,
\end{gather}
with $\beta=1\,{\rm km\,s}^{-1}$, a speed appropriate to disk ejections by Neptune-like planets. The orbits for each test particle were then integrated 30\,Myr forward to the present. We count the number of test particles in a 20\,pc box centered at the current position of the Sun to find a number density, $n=0.006\, {\rm pc}^{-3}$. Adjusting for the actual number of stars in CAR, we multiply by $(10/7)$ to get $n\sim0.0086\,{\rm pc}^{-3}$. In order to match \citet{2018ApJ...855L..10D}'s $\sim 0.2$ au$^{-3}$ estimate, of order $\sim 1\times 10^{21}$ `Oumuamua-like objects must be ejected from each star. Adopting a mass $M\sim 10^9$\,kg for 'Oumuamua \citep{2018ApJ...855L..10D}, this corresponds to $1.7\times 10^5\,M_{\Earth}$ or $\sim 540\,M_{\rm Jup}$, a value in line with the simpler estimate given above. 

A large number density is better understood if `Oumuamua's origin can be attributed to a formation process which is endemic to a molecular cloud core and which does not involve protostellar disks. A possible mechanism involving molecular hydrogen ice has been outlined by \citet{2020ApJ...896L...8S}, and is explored in depth by \citet{2020DPS....5231301L}, who conclude that H$_2$ ice deposition in molecular cloud cores is very difficult, but perhaps not impossible to achieve. Alternately,  \citet{2021arXiv210308812D} proposed that `Oumuamua is a nitrogen ice fragment produced by impacts onto Pluto-like nitrogen-surfaced objects in exoplanetary systems. This hypothesis satisfies the dynamical and photometric constraints, but requires a very high rate of such collisions.

In any event, we have shown that 'Oumuamua's galactic orbit is strongly suggestive of an origin in the COL or CAR moving group. Yet intriguingly, the large resulting inferred number density is in strongly conflict with a disk product hypothesis for its origin and is in significant tension with a cloud product hypothesis. Resolution of the mystery will very likely require observation of additional 'Oumuamua-like objects. 

\begin{acknowledgements}
      This material is based upon work supported by the National Aeronautics and Space Administration through the NASA Astrobiology Institute under Cooperative Agreement Notice NNH13ZDA017C issued through the Science Mission Directorate. We acknowledge support from the NASA Astrobiology Institute through a cooperative agreement between NASA Ames Research Center and Yale University.  The authors further thank Darryl and Sam for informative discussion on interstellar objects. 
\end{acknowledgements}

\software{Rebound, Astropy (The Astropy Collaboration 2013, 2018).}


\end{document}